\documentclass[prd,superscriptaddress,twocolumn,showpacs,amsmath,%
preprintnumbers,showkeys]{revtex4}
\usepackage{graphicx}
\usepackage{epstopdf}
\usepackage{epstopdf}
\usepackage{epsfig}
\def\beq{\begin{equation}}
\def\eeq{\end{equation}}

\def\beq{\begin{equation}}
  \def\eeq{\end{equation}}

\def\beq{\begin{equation}}
\def\eeq{\end{equation}}
\def\beeq{\begin{eqnarray}}
\def\eeeq{\end{eqnarray}}

\begin{document}
\title{
The four jet production at LHC and Tevatron in QCD. }
\author{ B. Blok\email{Email: blok@physics.technion.ac.il} }
\affiliation{Department of Physics, Technion---Israel Institute of Technology, 32000 Haifa, Israel}

\author{ Yu. Dokshitzer\email{Email: yuri@lpthe.jussieu.fr} }
\affiliation{Laboratory of Theoretical High Energy Physics (LPTHE), University Paris 6, Paris, France;
on leave of absence: PNPI, St.\ Petersburg, Russia}
\author{ L. Frankfurt\email{E-mail:
frankfur@tauphy.tau.ac.il} }
\affiliation{School of Physics and Astronomy, Raymond and Beverly
Sackler Faculty of Exact Sciences, Tel Aviv University, 69978 Tel
Aviv, Israel}
\author{ M.Strikman\email{\E-mail: strikman@phys.psu.edu}}
\affiliation{Physics Department, Penn State University, University Park, PA, USA}
\thispagestyle{empty}
\begin{abstract}
We demonstrate that in the back-to-back
 kinematics the production of four jets in the collision of two partons Êis suppressed
in the leading log approximation of pQCD,  compared to the Êhard
processes involving the collision of four partons. We derive the basic equation for 
four-jet production in QCD in terms of  the convolution of
generalized  two-parton distributions of colliding hadrons
in the momentum space representation.  Our derivation leads to geometrical approach
in the impact parameter space close to that suggested  within the parton model and used
before to describe the four-jet
production. We develop the independent parton approximation to the light-cone
wave function of the proton.
 Comparison with the CDF and D0 data shows that the independent parton approximation
 to the light-cone wave function of the proton is insufficient to explain the data. We argue that
 the data indicate the presence of significant
 multiparton correlations in the light-cone wave functions of colliding protons.
\end{abstract}

\maketitle
 \setcounter{page}{1}
\par
In spite of  extensive
 theoretical and experimental work,
various aspects  of the high-energy hadronic collisions at  the Tevatron and LHC are still poorly understood.
This is especially  true for the  multijet production which is of a paramount
importance for the understanding of pQCD dynamics at high-energy  colliders, and for the search of new particles.
  The topic of multiparton interactions is
now one of the focal points of studying pQCD and building an adequate basis for  modeling the final states at the LHC.
 In particular, description of multiparton interactions
requires treatment  of  a significant  inbalance of the momenta of
the jets  (presence of the Sudakov form factors).
 In this paper we summarize the first steps of the program to address these issues.  Among the original results of
 the paper are the derivation of the  formulas in the leading logarithmic  approximation for production of 4 jets.
 Our key finding is that it is possible to isolate the kinematics where the leading twist processes $2\to 4$ are not
  enhanced. This result will allow
one
to improve the reliability of the Tevatron studies of the four-jet production in the multiparton
kinematics and point out directions for the corresponding analysis at the LHC.

Another critical issue is the formulation of the
problem in terms of the generalized 
two-parton distributions
in the momentum space representation
and introduction of the mean field approximation for this
  object. This new formulation is well suited for the more detailed studies which are now under way.
   In addition it establishes a link with the original formulation in the coordinate space
\cite{Paver:1984ux,Mekhfi:1983az,PYTHIA, HERWIG, Treleani,Frankfurt, Frankfurtvi, SST, Wiedemann,Diehl:2010dr},
 and resolves an issue of the value of
the strength of the double interaction within this approximation. Previously there was a question
   whether a conclusion of Ref. \cite{Frankfurt},  that the observed rate is a factor of 2 larger
   than the theoretical prediction, can be due to
   uncertainties related to the many Fourier transforms  required to convert the HERA data
    to the experimental number.  A new formulation, though mathematically equivalent, has completely
     resolved this issue. This poses serious constraints on the Monte Carlo models of pp scattering at
      collider energies which are  not satisfied
by many of the current models.

These issues are of broad interest, both theoretical and experimental.
\par The standard approach to the multijet production is the QCD improved parton model. It is based on
the assumption  that the cross section of a
hard  hadron--hadron interaction
is calculable in terms of the convolution
of parton distributions within colliding hadrons
with the cross section of a hard two-parton collision.
An application of this approach to the processes
with production of four jets
implies that all jets in the event are  produced  in a
hard collision of {\em two}\/ initial state partons.

\par
The recent data of the  CDF and D0 Collaborations
\cite{Tevatron1,Tevatron2,Perugia} do not contradict to the
dominance of this mechanism in the well-defined part of the phase
space. At the same time these data provide the evidence that there
exists a kinematical domain where a more complicated mechanism
becomes important, namely the double hard interaction of two
partons in one hadron with two partons in the second  hadron.

\begin{figure}[t]  
   \centering
   \includegraphics[width=0.4\textwidth]{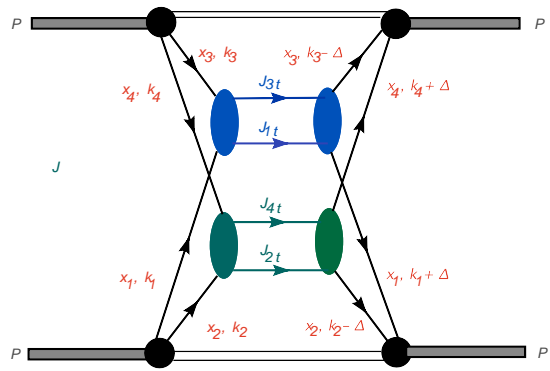}
   \caption{Kinematics of double hard collision - momenta of the colliding partons in $|in>$ and $<out|$ states.}
   \label{kin}
 \end{figure}

  Within the parton model picture, the four jets produced this way should pair into two groups such
   that the transverse momenta of two jets in each pair compensate each other.
  In what follows we refer to this kinematics as {\em back-to-back dijet production}.
  We consider the dijets for the case
\begin{equation}\label{eq:kinemo}
\delta_{13}^2\equiv (\vec{j}_{1t}+\vec{j}_{3t})^2 \ll \> j_{1t}^2\simeq j_{3t}^2, \>\>
\delta_{24}^2
 \ll \> j_{2t}^2\simeq j_{4t}^2,
\end{equation}
where $\delta$ is the total transverse momentum of the dijet and $j_{it}$ the transverse momentum of
an individual jet (see Fig.~\ref{kin}).
The hardness condition $\delta^2\gg R^{-2}$ is implied, with $R$ the characteristic
 hadron size (nonperturbative scale). The events with inbalances $\delta^2\le R^{-2}$ 
 give a small contribution both to
total and differential cross sections, since they are suppressed
by the Sudakov form factors. For a    detailed discussion of this
issue see   review \cite{DDT}.
\par
Importantly, in this kinematical region the hard scattering of four partons from the wave
functions of the colliding hadrons remains the dominant source for four-jet production even
 when the pQCD parton multiplication phenomena are taken into account.

The reason for that is the following. When the two partons from each hadron emerge from the
{\em initial state parton cascades}\/ and then engage into double hard scattering, the resulting differential
 distribution of the final state jets lacks the double back-to-back enhancement
 factor $d\sigma\propto \delta_{13}^{-2}\delta_{24}^{-2}$ which is there in the case
  of two independent hard scatterings~\cite{BDFS}.
For the two-parton scattering, the characteristic perturbative enhancement
  $d\sigma\propto \delta^{-2}$ results from a coherent enhancement of the
  amplitude due to integration over a large transverse disk, $\rho^2\sim \delta^{-2}\gg j_t^{-2}$.
  The two partons that originate from a perturbative splitting form a relatively compact system in
  the impact parameter space, so that the double hard interaction of such pairs produces only a
  single perturbative enhancement factor, $(\vec{\delta}_{13}+\vec{\delta}_{24})^{-2}$, which
  does not favor the back-to-back dijet kinematics \eqref{eq:kinemo}. The distribution of four
   jets so produced is much more isotropic and can be suppressed by choosing proper kinematical cuts.

\par So, the aim of this letter is to consider
the four-jet production in the hard collisions of {\em four}\/ initial state partons. We show that the cross section of back-to-back dijet production is calculable in terms of new
nonperturbative objects --- the generalized two parton distributions ($_2$GPD)
The properties of the
$_2$GPD
can be rigorously studied within QCD. In particular, we report here the derivation of the geometric picture for multiple  parton collisions in the impact parameter space.

In  the  kinematical domain \eqref{eq:kinemo} the direct calculation
of the light-cone Feynman diagrams (momenta of the  partons in the initial  and
final states
are shown in Fig.~\ref{kin})
using the separation of hard and soft scales shows that the four
$\to $ four cross section for the collisions of hadrons "a" and  "b" has the form:
\begin{eqnarray}\label{eq:main_form}
\sigma_4 (x_1,x_2,x_3,x_4)= &\int&
\frac{d^2\overrightarrow{\Delta}}{(2\pi)^2}
D_a(x_1,x_2,p_1^2,p_2^2,
\overrightarrow{\Delta})\nonumber\\[10pt]
\times D_b(x_3,x_4,p_1^2,p_2^2,-\overrightarrow{\Delta}&)&\times
\displaystyle{\frac{d\sigma^{13}}{d\hat t_1} \frac{d\sigma^{24}}{d\hat t_2}} d\hat t_1d\hat t_2.    
\label{b1}
\end{eqnarray}
Here $D_\alpha(x_1,x_2,p_1^2,p_2^2,  \overrightarrow{\Delta})$
are the new
$_2$GPDs
 for hadrons "a" and
"b" defined below.  (In the following we will consider the case of
$pp$ collisions and omit the subscripts $a$ and $b$. Summing over
collisions of various types of partons is implied.  In practice
however we will keep hard scattering of  gluons only since it
gives the dominant contribution.). Remember that the 
light-cone
fractions $x_i$ are actually fixed by the final jet parameters and
the energy momentum constraints.

With account of the radiative pQCD effects, in full analogy with the "DDT formula" for two-body collisions,
the differential distribution \eqref{eq:main_form} acquires Sudakov form factors~\cite{DDT,Catani} depending
on the logarithms of the large ratios of scales, $j_t^2/\delta^2$, and the $_2$GPDs become scale dependent:
$p_1^2\sim \delta_{13}^2$, $p_2^2\sim\delta_{24}^2$.  It should be mentioned that the structure of the final
formula depends on what one actually measures in the experiment ---   energetic single particles with
large transverse momenta in the final state or "jets" --- and on how the jets are precisely defined.
A more detailed account of the pQCD effects will be given in a future publication~\cite{BDFS}.

For brevity we will not write explicitly the virtuality scales of the
$_2$GPD
and will use the form:
$D(x_1,x_2,\overrightarrow{\Delta} )$. Note that these distributions depend on the new transverse vector  $\overrightarrow{\Delta}$ that
is equal to the difference of the momenta of partons from the wave function of the colliding hadron
in the amplitude and the amplitude conjugated.
Such dependence arises because the difference of parton transverse momenta within
the parton pair is not conserved. The integration limits in $x_i, \hat t$ are subject to
standard limits determined by experimental kinematic cuts.

\par
Within the parton model approximation the cross section has the form:
 \beq
 \sigma_4= \sigma_1\sigma_2/\pi R^2_{\rm int},
 \label{b2}
 \eeq
where $\sigma_1$ and $\sigma_2$ are the cross sections of two independent hard binary parton interactions. The factor $\pi R^2_{\rm int}$ which in principle depends on $x_i$ characterizes the transverse area occupied by the
partons participating in the hard collision. (In  the experimental  \cite{Tevatron1,Tevatron2}  and some of the theoretical  papers this factor was denoted  as an effective cross section. Our Eq.~\ref{b3} below  shows that such wording  is not satisfactory   since
$\pi R^2_{\rm int}$ does not have the  meaning of the interaction cross section.) The data
\cite{Tevatron1,Tevatron2,Perugia} indicate
that $\pi R^2_{\rm int}$ is practically constant in the kinematical range studied at the Tevatron.

 Eq.~\ref{b1} leads to  the general model
independent expression for
 \beq
\frac{1}{\pi R^2_{\rm int}}=\int
 \frac{d^2\overrightarrow{\Delta}}{(2\pi)^2}\frac{D(x_1,x_2,-\overrightarrow{\Delta})
 D(x_3,x_4,\overrightarrow{\Delta})}{D(x_1)D(x_2)D(x_3)D(x_4)},
 \label{b3}
 \eeq
 in terms of  the $_2$GPDs.
 Here $D(x_i)$ are the
 corresponding structure functions. Note that hereafter we do not
 write dependence of $\sigma_4$ and $R^2_{\rm int}$ on the light-cone fractions $x_i$ explicitly.

\par The
$_2$GPDs
are expressed through the light-cone wave functions of the colliding  hadrons as follows. Suppose that in a four $\to$ four
process  the two partons in the nucleon in the initial state wave
function have the transverse momenta
$\overrightarrow{k_1},\overrightarrow{k_2}$. Then  in the
conjugated wave function they will have the momenta
$\overrightarrow{k_1}+\overrightarrow{\Delta},\overrightarrow{k_2}-\overrightarrow{\Delta}$.
This is because only the sum of parton transverse momenta but not the
difference is  conserved.

The relevant
$_2$GPDs
are:
\begin{eqnarray}
 &D&(x_1,x_2,p^2_1,p^2_2,\overrightarrow{\Delta})=\sum_{n=3}^{\infty}\int
\frac{d^2k_1}{(2\pi)^2}\frac{d^2k_2}{(2\pi)^2}\theta
(p_1^2-k_1^2)\nonumber\\[10pt]
&\times& \theta (p_2^2-k_2^2)\int \prod_{i\ne
1,2}\frac{d^2k_i}{(2\pi)^2}\int^1_0\prod_{i\ne
1,2} dx_i\nonumber\\[10pt]
&\times& 
\psi_n (x_1,\vec k_1,x_2,\vec k_2,.,\vec k_i,x_i..)
\nonumber\\[10pt]
&\times&\psi_n^+(x_1,\overrightarrow{k_1}+\overrightarrow{\Delta},x_2,\overrightarrow{k_2}
-\overrightarrow{\Delta},x_3, \vec k_3,...) 
\nonumber\\[10pt]
&\times&  (2\pi)^3\delta( \sum_{i=1}^{i_=n} x_i-1)\delta (\sum_{i=1}^{i=n} \vec
k_i).
\label{b4}
\end{eqnarray}
Note that this distribution is diagonal in the space of all
partons except the two partons involved
 in the collision. Here $\psi$ is the parton
wave function normalized to one in a usual way. An appropriate
summation over color and Lorentz indices is implied. In the case
of kinematics  $1\gg x_1\ge x_2$ we expect only distributions
without the spin flip  to be important. \par Let us stress that it
follows from the above formulas that in the impact parameter space  these GPDs have a
 probabilistic interpretation. In particular  they are positively definite  in the impact parameter space, cf.\ Eq.~\ref{l1}.  Note that in the same way one can introduce the $N$-particle GPD, $G_N$,
which can be probed in the production of $N$ pairs of jets.  In this case  the   first $N$ arguments $k_i$ in Eq.~\ref{b4}  are shifted by
$\overrightarrow{\Delta_i}$ subject to the  constraint $\sum_i \overrightarrow{\Delta_i}=0$. So the cross
section is proportional to
\begin{eqnarray}
\sigma_{2N}&\propto& \int \prod_{i=1}^{i=N}{d\overrightarrow{\Delta}_i\over (2\pi)^2}
D_a(\overrightarrow{\Delta}_1,...\overrightarrow{\Delta}_N)  \nonumber\\[10pt]
 &\times& D_b(\overrightarrow{\Delta}_1,...\overrightarrow{\Delta}_N)
\delta(\sum_{i=1}^{i=N}\overrightarrow{\Delta}_i).
\end{eqnarray}

\par These GPDs can be easily rewritten in the form of the matrix elements  of
the operator product.
For example:
\begin{eqnarray}  D(\Delta )&=& <N\vert \int
d^4x_1d^4x_2d^4x_3  \nonumber\\[10pt]
&\times &G^a_{i+}(x_1)G^b_{j+}(x_2)G^a_{i+}(x_3)G^b_{j+}(x_4)\nonumber\\[10pt]
&\times&\exp(ip_1^+(x_1-x_3)^-+ip_2^+(x_2-x_4)^-  \nonumber\\[10pt]
&+&
 i\vec\Delta_t (\vec{x}_4-\vec{x}_3)_t) \vert N>,
\label{1b}\end{eqnarray} calculated at the virtualities
$p_1^2,p_2^2$ at fixed $\overrightarrow{\Delta}$. Here we gave an
example for the most relevant case of gluons without a flip in
color and spin spaces. In general a number of distributions can be
written, depending on different contractions of transverse Lorentz
indices and color indices. The classification of the relevant
distributions is the same as the classification of the
quasipartonic operators in Ref. \cite{Lipatov}. Note that the
presence of the transverse external parameter $\vec\Delta$ does
not change the classification, since the corresponding new
structures will be strongly suppressed at high energies. We wrote
the operator expression in the light-cone gauge. In an arbitrary  gauge we shall need Wilson
lines
W(C) connecting points with contracted color indices.
\par In the approximation of
uncorrelated
partons it follows from  Eq.~\ref{b4} that
\begin{equation}D(x_1,x_2,p_1^2, p_2^2, \vec\Delta
)= G(x_1, p_1^2, \vec\Delta)G(x_2, p_2^2, \vec\Delta ),\end{equation}
 where
$G(x,\overrightarrow{\Delta} )$ are conventional one-particle
GPDs. These GPDs can be approximated as  $G_N(x,Q^2, \vec \Delta)=
G_N(x,Q^2) F_{2g}(\Delta)$, where $F_{2g}(\Delta)$ is the two-gluon
form factor of
 the nucleon extracted from hard exclusive vector meson
 production
  (we suppress here the dependence of $ F_{2g}$ on x)
  \cite{FS} and $G_N(x,Q^2)$ conventional parton distribution of a nucleon.
(Here $Q^2$ is the virtuality due to the radiation, cf.\ discussion
after
 Eq.~\ref{b1}.) Thus :
\beq \frac{1}{\pi R^2_{\rm int}}=\int
\frac{d^2\Delta}{(2\pi)^2}F_{2g}^4(\Delta)=\frac{m^2_g}{28\pi}.\label{b6}\eeq
Here at the last step we used the dipole fit
$F_{2g}(\Delta)=1/(\Delta^2/m^2_g+1)^2$ to the two-gluon form
factor ($m^2_g(x\sim 0.03, Q^2= 3\mbox{GeV}^2)\approx 1.1 \mbox{GeV}^2$). Using the transverse gluon radius of the nucleon we obtain
\beq R^2_{\rm int}=7/2r^2_g,~~~~~r^2_g/4=d F_{2g}(t)/dt_{t=0}.
\label{b7} \eeq
 This result coincides with the one for the area $\pi R^2_{\rm int}$ obtained earlier in \cite{Frankfurt}
 using the geometric picture in the impact parameter space.
That derivation
involved
taking  the Fourier transform of the two-gluon form factor and calculating  a rather
complicated six-dimensional integral which could potentially lead to large numerical uncertainties.
The form of Eq.~\ref{b7} clearly indicates that numerical
uncertainties are small.
\par It was emphasized in \cite{Frankfurt} that the experiments on
four-jet production  report a smaller value of $\pi R^2_{\rm int}$
as compared to the one obtained above in the independent particle
approximation (though the issue of how well the contribution of the $2\to 4$ processes was  subtracted
 still remains, cf. discussion in beginning of the paper). It is at least a factor of 2  smaller ---
 that is a four-jet cross section is a factor of 2 larger ---
than  Eq.~\ref{b7} gives. (The GPDs for sea quarks appear to decrease
with $\Delta$ somewhat faster, resulting in a smaller $1/\pi R^2_{int}$, see discussion in \cite{Strikman:2009bd}.)
\par  It  follows from Eq.~\ref{b3} that the value of
$R^2_{\rm int}$ is determined by the range of integration over
$\Delta$. Hence the characteristic  $\Delta $ in the integral
measures the  effective distance between the parton pairs (which
in principle may differ for different flavor combinations).
 According to the above
evaluation within the independent parton approximation the
integral for $1/R^2_{\rm int}$ is dominated by small
$\Delta^2\sim 0.1 \,m^2_g$.
 The contribution of large $\Delta$ is suppressed
by the two-gluon form factor of a nucleon. This reasoning indicates
the important role of interparton correlations.  In other words,
the integral
 over $\Delta$ is effectively cut off  by a scale of the nonperturbative correlations.
  Such correlations
 naturally arise in nonperturbative QCD regime
in a number of nucleon models, such as constituent quark model
(gluon cloud around constituent quark) \cite{Frankfurt}, or
string model (gluon structure of string)   \cite{Bj}.
The detailed analysis of the additional correlations due to the hard-- soft interplay will be reported
 elsewhere \cite{BDFS}.

 Let us now show that results obtained in the paper lead to the  geometric picture
in the impact parameter space mentioned above
 \cite{Paver:1984ux,Mekhfi:1983az,PYTHIA, HERWIG,Treleani, Frankfurt, Frankfurtvi, SST, Wiedemann,Diehl:2010dr}.

 The first step is to make transformation into coordinate space  i.e., to make the Fourier transform from variables $k_i$ in  Eq.~\ref{b4} to coordinates
$\rho_i$. Performing integration over  $k_i$ we obtain that transverse coordinates of partons  in the
amplitude and the amplitude conjugated are equal
$\rho_i=\rho_f$.
 In the calculation we use the fact  that upper limit of integration over $k^2_t$
is very large compared with the inverse hadron size. The next step is to perform integration over $\Delta$  which produces
$\delta(\vec \rho_1-\vec\rho_2-\vec \rho_3+\vec \rho_4)=\int d^2B \delta (\vec \rho_1-\vec \rho_3-\vec B)
 \delta(\vec \rho_2-\vec \rho_4-\vec B)$.
\par  The delta functions express the fact that within the accuracy $1/p_t$ where $p_t$ is the hard scale,
the interactions of partons  from different nucleons occur at the same point.  $\vec  B$ is
the relative impact parameter of two  nucleons.
\par
\begin{figure}[h]  
   \centering
   \includegraphics[width=0.3\textwidth]{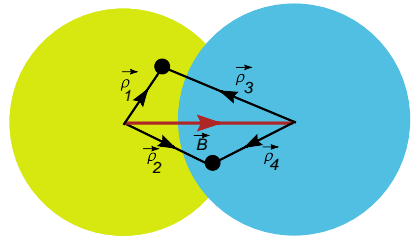}
   \caption{Geometry of two hard collisions in impact parameter picture.}
    \label{bdistr}
 \end{figure}

The expression for the cross  section in the impact parameter  space has the form
 which corresponds to the geometry of Fig.\ref{bdistr}
\begin{eqnarray}
\sigma_4&=&\int d^2B
d^2\rho_1d^2\rho_2d^2\rho_3d^2\rho_4D(x_1,x_2,\vec \rho_1,\vec \rho_2)\nonumber\\[10pt]
&\times&D(x_3,x_4,\vec \rho_3,\vec \rho_4)
\delta (\vec \rho_1-(\vec B+\vec \rho_3))\delta (\vec \rho_2-(\vec B+\vec \rho_4))=\nonumber\\[10pt]
& = &\int d^2B
d^2\rho_1d^2\rho_2 D(x_1,x_2,\vec \rho_1,\vec \rho_2)  \nonumber\\[10pt]
&\times& D(x_3,x_4,-\vec B+\vec \rho_1,-\vec B+\vec \rho_2). 
\label{l1} \end{eqnarray} Here the
$_2$GPD
in the impact parameter space representation  is given by
\begin{eqnarray}
&D&(x_1,x_2,\vec \rho_1,\vec \rho_2)= \nonumber\\[10pt]
& = & \sum^{n=\infty}_{n=3}\int \prod_{i\ge 3}^{i=n}\left[dx_i d^2\rho_i\right] \psi_n(x_1,\vec \rho_1,x_2,\vec \rho_2,...x_i,\vec \rho_i,) \nonumber\\[10pt]
&\times &\psi_n^+(x_1,\vec \rho_1,x_2,\vec \rho_2,...,x_i,\vec \rho_i,...)\delta(\sum_{i=1}^{i=n}  x_i\vec\rho_i).
\label{o3}
\end{eqnarray}
where the delta function expresses the center of mass constraint
$\sum_{i=1}^{i=n} x_i\vec\rho_i
=0 $. This is analogous to the case of single parton GPDs, see  \cite{Diehl1}.
 The functions
$\psi (x_1,\vec \rho_1,x_2,\vec \rho_2,...)$ are just the Fourier
transforms in the impact parameter space of the light-cone wave
functions and are given by
\begin{eqnarray}
&\psi_n& (x_1,\vec \rho_1,x_2,\vec \rho_2,...)=\int \prod^{i=n}_{i=1}\frac{d^2k_i}{(2\pi)^2}\exp(i\sum_{i=1}^{i=n}\vec k_i\vec \rho_i)\nonumber\\[10pt]
&\times&\psi_n(x_1,\vec
k_1,x_2,\vec k_2,..)(2\pi)^2
\delta (\sum \vec k_i).\label{o4}
\end{eqnarray}

\par Thus the $_2$GPD based description of the $four \to four$ processes is   equivalent to  the representation for the cross
 section  corresponding  to the
simple geometrical picture, but instead of a triple integral we now  have  an  integral over
one  momentum  $\Delta$.
The $_2$GPD defined in  Eq.~\ref{b4} is useful for calculation of many different processes.
At the same time the knowledge of the full double GPD is necessary for complete description of
 events with a double jet trigger since the pedestal strongly depends on the impact parameter $\vec B$ \cite{Frankfurt}.

\par Let us stress that this picture is a natural generalization of the
correspondence between momentum representation and geometric
picture for a conventional case of two $\to$ two collisions. Indeed in this case it is easy to
 see that the cross section in the
momentum representation \beq \sigma_2=\int
f(x_1,p^2)f(x_2,p^2)\frac{d\sigma^h}{d\hat t}d\hat t\label{s1}\eeq
has a simple geometric representation
\beq \sigma_2=\int
d^2\rho_1d^2B f(x_1,\vec \rho_1,p^2)f(x_2,\vec
B-\vec{\rho}_1,p^2)\frac{d\sigma^h}{d\hat t}d\hat t,\label{s2}\eeq
 where
$f(x,\vec \rho,p^2)= \psi^+(x,\vec \rho,p^2)\psi(x,\vec \rho,p^2)$
and $\psi(x, \vec \rho,p^2)$ is the Fourier transform of the light-cone
wave function defined above.
\par Let us now summarize our results. We have argued that there exists
the kinematical domain  where the four $\to$ four hard parton
collisions form the dominant mechanism of four-jet production. In
this region we calculated the cross section, see
Eqs.~\ref{b1}-\ref{b3}  and found that it can be expressed through
new  $_2$GPDs (see Eq.~\ref{b4}), expressed through the light-cone
wave functions of the colliding hadrons. These $_2$GPDs depend on
a transverse vector $\vec \Delta$ that measures the transverse
distance  within the parton pairs. (Equivalent expressions for
these GPDs can be easily  given in terms of the operator
products.) In the impact parameter space we derived the widely
used intuitive geometric picture. We argued   that  the observed
enhancement of a four-jet cross section indicates  the presence of
short-range two-parton correlations in the nucleon parton wave
function, as determined by the range of integral over  $\Delta$.
The contribution of perturbative correlations in the appropriate
kinematic domain is suppressed. The detailed study of the
interplay of the contribution of hard/soft correlations will be
reported elsewhere \cite{BDFS}.
\par It was argued  recently  \cite{Stirling} that the cross
section can be expressed in terms of two-parton distribution
functions. Our analysis indicates that  a more detailed treatment
of the QCD evolution effects   is necessary. We found that it is
necessary to introduce the new 2-particle $_2$GPDs
which depend on
additional parameter $\Delta$.  The parameter $\Delta$ expresses
the fact that the difference of the  transverse components of the
parton momenta is not conserved and therefore different in $\left|
in \right>$ and $\left<out\right|$ states in the double hard
collisions.

\section*{Acknowledgements}
Two  of us (LF and MS) would like to thank the Yukawa International
Program for Quark--Hadron Sciences  for hospitality during
a part  of this study.
This research was supported by the
United States Department of Energy and the Binational Science Foundation.


\begin{thebibliography}{}


\bibitem{Paver:1984ux}
  N.~Paver and D.~Treleani,
  Z.\ Phys.\  C {\bf 28}, 187 (1985).
\bibitem{Mekhfi:1983az}
  M.~Mekhfi,
  Phys.\ Rev.\  D {\bf 32}, 2371 (1985).


\bibitem{Treleani}A.~Del Fabbro and D.~Treleani,
  Phys.\ Rev.\  D {\bf 61}, 077502 (2000)
  [arXiv:hep-ph/9911358]; Phys.\ Rev.\  D {\bf 63}, 057901 (2001)
  [arXiv:hep-ph/0005273];  A.~Accardi and D.~Treleani,
  Phys.\ Rev.\  D {\bf 63}, 116002 (2001)
  [arXiv:hep-ph/0009234].
  \bibitem{Frankfurt}
  L.~Frankfurt, M.~Strikman and C.~Weiss,
  Phys.\ Rev.\  D {\bf 69}, 114010 (2004)
  [arXiv:hep-ph/0311231],
  Ann.\ Rev.\ Nucl.\ Part.\ Sci.\  {\bf 55}, 403 (2005)
  [arXiv:hep-ph/0507286].

\bibitem{PYTHIA} for a recent summary see  T.~Sjostrand and P.~Z.~Skands,
  Eur.\ Phys.\ J.\  C {\bf 39}, 129 (2005)
  [arXiv:hep-ph/0408302].
\bibitem{HERWIG}for a recent summary see   M.~Bahr {\it et al.},
  Eur.\ Phys.\ J.\  C {\bf 58}, 639 (2008)
  [arXiv:0803.0883 [hep-ph]].

\bibitem{Wiedemann} S.~Domdey, H.~J.~Pirner and U.~A.~Wiedemann,
  Eur.\ Phys.\ J.\  C {\bf 65}, 153 (2010)
  [arXiv:0906.4335 [hep-ph]].

\bibitem{Frankfurtvi}
  L.~Frankfurt, M.~Strikman, D.~Treleani and C.~Weiss,
  Phys.\ Rev.\ Lett.\  {\bf 101}, 202003 (2008)
  [arXiv:0808.0182 [hep-ph]].
\bibitem{SST}  T.~C.~Rogers, A.~M.~Stasto and M.~I.~Strikman,
  Phys.\ Rev.\  D {\bf 77}, 114009 (2008)
  [arXiv:0801.0303 [hep-ph]].
  \bibitem{Diehl:2010dr}
  M.~Diehl,
  arXiv:1007.5477 [hep-ph].
\bibitem{Tevatron1}F.~Abe {\it et al.}  [CDF Collaboration],
  Phys.\ Rev.\  D {\bf 56}, 3811 (1997).

\bibitem{Tevatron2}  V.~M.~Abazov {\it et al.}  [D0 Collaboration],
  Phys.\ Rev.\  D {\bf 81}, 052012 (2010)
  [arXiv:0912.5104 [hep-ex]].
\bibitem{Perugia} Proceedings of the 1st International Workshop on  Multiple Partonic Interactions at the LHC,
Verlug Deutshes electronen synchrotron, 2010.

 \bibitem{DDT}
  Y.~L.~Dokshitzer, D.~Diakonov and S.~I.~Troian,
  Phys.\ Rept.\  {\bf 58}, 269 (1980).
  
\bibitem{BDFS} B. Blok, Yu. Dokshitzer, L. Frankfurt, M. Strikman
  unpublished

\bibitem{Catani}
  S.~Catani, E.~D'Emilio and L.~Trentadue,
  Phys.\ Lett.\  B {\bf 211}, 335 (1988).



\bibitem{Lipatov} A.P. Bukhvostov, G.V.~ Frolov, L.N.~ Lipatov,~ E.A. Kuraev, Nucl.\ Phys.{\bf B258}, 601 (1985).
\bibitem{FS} L.~Frankfurt and M.~Strikman,
  Phys.\ Rev.\  D {\bf 66}, 031502 (2002)
  [arXiv:hep-ph/0205223].
  \bibitem{Strikman:2009bd}
  M.~Strikman and C.~Weiss,
  Phys.\ Rev.\  D {\bf 80}, 114029 (2009)
  [arXiv:0906.3267 [hep-ph]].

\bibitem{Bj} J.~Bjorken, a talk at the Spring workshop on electron-Nucleus Collider Physics, May 14, 2010.
\bibitem{Diehl1}
 M.~Diehl,
  Eur.\ Phys.\ J.\  C {\bf 25} (2002) 223
  [Erratum-ibid.\  C {\bf 31} (2003) 277]
  [arXiv:hep-ph/0205208];
M.~Diehl,
  Phys.\ Rept.\  {\bf 388}, 41 (2003)
  [arXiv:hep-ph/0307382].


\bibitem{Stirling}  J.~R.~Gaunt and W.~J.~Stirling,
  JHEP {\bf 1003}, 005 (2010)
  [arXiv:0910.4347 [hep-ph]];
  J.~R.~Gaunt, C.~H.~Kom, A.~Kulesza and W.~J.~Stirling,
  Eur.\ Phys.\ J.\  C {\bf 69}, 53 (2010)
  [arXiv:1003.3953 [hep-ph]]
\end{thebibliography}
\end{document}